\documentclass[preprint]{aastex63}
\usepackage{amsmath}
\usepackage{multirow}
\usepackage{sidecap}
\usepackage{floatrow}

\received{November 6, 2019}
\revised{March 5, 2020}
\accepted{TBD}
\submitjournal{AJ}

\shorttitle{Nodal Precession}
\shortauthors{Bailey \& Fabrycky}
\graphicspath{{./}{figures/}}

\begin{document}

\title{Nodal Precession in Closely Spaced Planet Pairs}

\correspondingauthor{Nora Bailey}
\email{norabailey@uchicago.edu}

\author[0000-0001-7509-0563]{Nora Bailey}
\affiliation{Department of Astronomy \& Astrophysics, University of Chicago, Chicago, IL 60637}

\author[0000-0003-3750-0183]{Daniel Fabrycky}
\affiliation{Department of Astronomy \& Astrophysics, University of Chicago, Chicago, IL 60637}

\begin{abstract}

Planet-planet perturbations can cause planets' orbital elements to change on secular timescales. Previous work has evaluated the nodal precession rate for planets in the limit of low $\alpha$ (semi-major axis ratio, 0$<$$\alpha$$\leq$1). Our simulations show that systems at high $\alpha$ (or low period ratio), similar to multiplanet systems found in the \emph{Kepler} survey, have a nodal precession rate that is more strongly dependent on eccentricity and inclination. We present a complete expansion of the nodal precession rate to fourth order in the disturbing function and show that this analytical solution much better describes the simulated N-body behavior of high-$\alpha$ planet pairs; at $\alpha\approx$ 0.5, the fourth-order solution on average reduces the median analytical error by a factor of 7.5 from linear theory and 6.2 from a second-order expansion. We set limits on eccentricity and inclination where the theory is precisely validated by N-body integrations, which can be useful in future secular treatments of planetary systems.

\end{abstract}


\section{Introduction} \label{sec:intro}

When multiple planets orbit a star, the planets interact with one another and thus their orbits change over time. Systems with multiple planets are very common--aside from our own solar system with its 8 planets, 1735 of the confirmed 4084 exoplanets reside in multiplanet systems (NASA Exoplanet Archive\footnote{https://exoplanetarchive.ipac.caltech.edu, as of 28 October 2019 }), and it's likely that most of the single-planet systems have undetected companions. Understanding how planets interact is necessary to understanding the architectures, stability, and long-term behavior of exoplanet systems.

Laplace-Lagrange theory tells us how orbital parameters change over time for planets using linear secular theory. A pair of planets has one nonzero inclination-node eigenfrequency, given by solving the characteristic equation for the Laplace-Lagrange secular solution as shown in \citet[Equation 7.31]{MurrayDermott}. This frequency is equivalent to the planet's change in ascending node per time ($\dot{\Omega}$). An expression for $\dot{\Omega}$ following this derivation is

\begin{equation}\label{eqn:gll}
    \dot{\Omega}_{\text{LL}}=-\frac{1}{4}b_{3/2}^{(1)}(\alpha)\alpha\left ( n_1\frac{m_2}{M_\star+m_1} \alpha + n_2\frac{m_1}{M_\star+m_2} \right ).
\end{equation}

In this equation, for planets 1 and 2, $\alpha$ is the ratio of semi-major axes ($a_1/a_2$, $a_1<a_2$), $n_{1,2}$ are the mean motions, and $b_{3/2}^{(1)}(\alpha)$ is the Laplace coefficient (see Equation~\ref{eqn:laplacecoeff} for the general definition). $\dot{\Omega}$ has the same units as the mean motions and is equal for both planets. All orbital elements are given relative to the invariable plane and the central star.

\cite{2011Lithwick} examine the eccentricity and inclination dependence of the nodal precession. They find an expression for $\dot{\Omega}$ for a test particle interior to a circular massive planet, neglecting terms in $e^2i^2$, $e^4$, $i^4$ and higher, in their Equation 33, reproduced here:

\begin{equation}\label{eqn:LW11eqn33}
    \dot{\Omega} = -\gamma(1-\frac{1}{2}i^2+2e^2)
\end{equation}

Their $\gamma$ is equivalent to $\dot{\Omega}_{\text{LL}}$ from Equation~\ref{eqn:gll} (assuming $m_1$=0) except that \cite{2011Lithwick} use the approximation that $b_{3/2}^{(1)}(\alpha)$ = 3$\alpha$ for $\alpha\ll1$ \citep{2007Heyl}. Additionally, for the case where the test particle is \textit{exterior} to a massive planet ($m_2$=0), the $\gamma$ expression would need to be modified with a factor of 1/$\alpha$ (this difference between the inner and outer particles can be seen in the two terms of Equation~\ref{eqn:gll}).

The focus of \cite{2011Lithwick} is on the solar system, where the $\alpha$ values are low (particularly between terrestrial planets and Jupiter), and their analysis only includes the first order of $\alpha$ in the Hamiltonian. In exoplanet systems, however, $\alpha$ is frequently larger, and the effect of $\alpha$ on the nodal precession rate may be non-negligible. The goal of this work is to find how the nodal precession rate varies with inclination, eccentricity, and semi-major axis ratio, particularly for the values relevant for packed (though non-resonant) multiplanetary systems like those found by \emph{Kepler}.

Previous work has largely focused on coplanar systems, both to simplify the problem and as a reasonable approximation given that the mutual inclination of exoplanet systems is generally either small \citep{2014Fabrycky} or unknown. Coplanar studies have attempted to describe the secular eccentricity evolution of exoplanets systems, finding that fourth-order expansions are not sufficient \citep{2007Veras} but twelfth-order expansions can succeed \citep{2006Libert}. Similarly, it is to be expected that higher order terms will be needed to describe non-coplanar secular dynamics, as has been demonstrated in \cite{2008Libert} for 3-D secular frequencies modeling the $\upsilon$ Andromedae planetary system with a mutual inclination $\sim 20^\circ$. Additionally, studies that have included inclination effects (e.g. \citealt{2019Volpi}) tend to be focused on the stability and dynamical effects and not on nodal precession itself.

Secular nodal precession occurs on timescales unlikely to affect observable orbits, but the secular nodal precession frequency can have dynamical implications, particularly via secular resonances. A secular resonance occurs when a linear combination of the apsidal and/or nodal precession frequencies are integer multiples of one another. Because the location of the inclination-node secular resonance depends on the nodal precession frequency, it is potentially sensitive to higher-order effects.

Further, the overlap of these secular resonances can lead to chaos and instabilities in a planetary system. Both \cite{2011Lithwick} and \cite{2012Boue} have shown that, in the case of Mercury's excitation, the solar system's few-degree mutual inclinations play an important role in the system's transition to chaos. The location and width of the secular resonances depends on the nonlinear deviations from Laplace-Lagrange theory, and high-$\alpha$ effects could have implications for the onset of chaos in exoplanet systems even at mild inclinations and eccentricities.

In this work, we use both numerical and analytical methods to examine how the nodal precession rate varies with $e$, $i$, and $\alpha$. Our analytical approach here is to expand the disturbing function to fourth order in $e$ and $i$, keeping all associated higher orders of $\alpha$. This approach contrasts with other approaches in the literature, such as \cite{2010LaskarBoue} in which $e$ and $i$ are arbitrary and the expansion of the disturbing function is in $\alpha$ as a small parameter.

In Section~\ref{sec:numinv}, we describe simulations of restricted and unrestricted 2-planet systems and their nodal precession compared with linear theory. In Section~\ref{sec:analytical}, we expand the nodal precession rate from \cite{MurrayDermott} to fourth order in $e$ and $i$, keeping all orders of $\alpha$, and compare this to our numerical results. In Section~\ref{sec:applicablerange}, we investigate the range of parameter space where our analytical solution is accurate. In Section~\ref{sec:Kepler}, we discuss the implications for multiplanet systems similar to those discovered by \emph{Kepler}. In Section~\ref{sec:secularres}, we show how an example of how the location of the secular resonance could be affected using the system \emph{Kepler}-117. Finally, we discuss our conclusions in Section~\ref{sec:concl}.

\section{Numerical Investigation}\label{sec:numinv}

\subsection{Restricted Simulations}\label{sec:restsims}

To examine the effect of $\alpha$ on nodal precession, we ran a suite of restricted three-body simulations with a massive circular planet and a test particle. There were two complete sets of simulations, one with the test particle as the interior body and one with the test particle as the outer body. The outcomes of both were similar, so here we will present only the case of the massive planet interior to an outer test particle.

\subsubsection{Initial Conditions and Integration}

Period ratio and $\alpha$ describe the same property of the planet pair--their relative spacing--and are related via Kepler's Third Law: 

\begin{equation}\label{eqn:alphaPR}
    \alpha^3 = \frac{m_1+M_\star}{(m_2+M_\star)(P_2/P_1)^2}
\end{equation}

We choose to use period ratio as our parameter space here as it is the directly measurable quantity and for the ease in avoiding mean motion resonances (MMRs), where the planets' periods are integer multiples of one another.

The initial values for the system were randomly generated and then used consistently for each simulation. The inner planet's mass was chosen uniformly between 2 and 10 $M_\oplus$. The inner planet's period was chosen uniformly between 3 and 6 days. The outer planet's longitude of ascending node ($\Omega$) and argument of pericenter ($\omega$) were chosen uniformly between -$\pi$ and $\pi$.

The period, eccentricity, and inclination of the outer test particle were varied across all 3 parameters to cover a region up to 50 in period ratio (avoiding first- and second-order MMRs by $\pm$0.1 such that $|P_2/P_1-M/N| \geq 0.1$ for $M=N+1$ and $M=N+2$, which results in a lower bound of 1.77 in period ratio), from 0 to 0.1 in eccentricity, and from 0.5 to 6 degrees in inclination. The corresponding range of $\alpha$ is 0.0737 to 0.6834.

The values of $\Omega$ were saved every 0.4$P_1$, and each system was integrated for 10$^5P_2$, allowing to observe the majority of a nodal precession cycle for even the highest period ratio. Initial system parameters are shown in Table~\ref{tab:simICs}. The integrations were computed using the \texttt{REBOUND} package with the IAS15 integrator \citep{2012Rein,2015Rein1}.

\begin{table}[h!]
\centering
\caption{Initial conditions for the restricted simulations with outer test particle.}
\label{tab:simICs}
\begin{tabular}{|l|l|l|}
\hline
Stellar Mass & \multicolumn{2}{l|}{0.5 $M_\odot$} \\ \hline
\textbf{Planet:} & \textbf{b} & \textbf{c} \\ \hline
Period (days) & 3.4834 & ($P_2/P_1$)$\times$3.4834 \\ \hline
Mass & 7.3336 $M_\oplus$ & 0 \\ \hline
Eccentricity & 0 & $e$ \\ \hline
Inclination & 0 & $i$ \\ \hline
$\Omega$ & N/A & 1.0519 rad \\ \hline
$\omega$ & N/A & 1.0944 rad \\ \hline
\textbf{Variable Parameters} & \textbf{Min} & \textbf{Max} \\ \hline
$P_2/P_1$ & 1.77 & 50 \\ \hline
$e$ & 0 & 0.1 \\ \hline
$i$ (deg) & 0.5 & 6 \\ \hline
\end{tabular}
\end{table}

\subsubsection{Simulation Analysis}\label{sec:simanal}

To analyze the output from the simulations, we first checked if the system was stable in period ratio and did not vary more than 0.03 from the starting period ratio. If such a departure occurred, only data from the initial time period before the departure was used for the analysis. Applying this check allowed the initial period ratio to be considered characteristic for the system as well as ensuring the secular approximation that semi-major axis and thus $\alpha$ are constant over time is valid.

Then from the values of $\Omega$ over time for each simulation, we fit a sinusoid to cos$\,\Omega$ to determine the average period of nodal precession. Typical error bars associated with the period are smaller than the plotted points, although there is some time variation in the nodal precession (see Section~\ref{sec:timedepend}). This method takes advantage of the fact that there is a single expected frequency for the pair and is faster and more accurate than more complicated methods, such as Fourier analysis.

We also calculated a predicted period from Laplace-Lagrange theory for each simulation as $P_{\text{LL}}=\frac{2\pi}{|\dot{\Omega}_{\text{LL}}|}$ to be used to scale the periods for uniform comparison.

\subsubsection{Results}\label{sec:numres_rest}

When the observed simulated nodal precession periods were compared to linear theory, the results were stark. Deviations were seen as high as 12\% for relatively low values of inclination and eccentricity, with a very sensitive dependence on $\alpha$. The changes were much greater than that predicted by Equation~\ref{eqn:LW11eqn33}.

\begin{figure*}[h]
  \includegraphics[width=\linewidth]{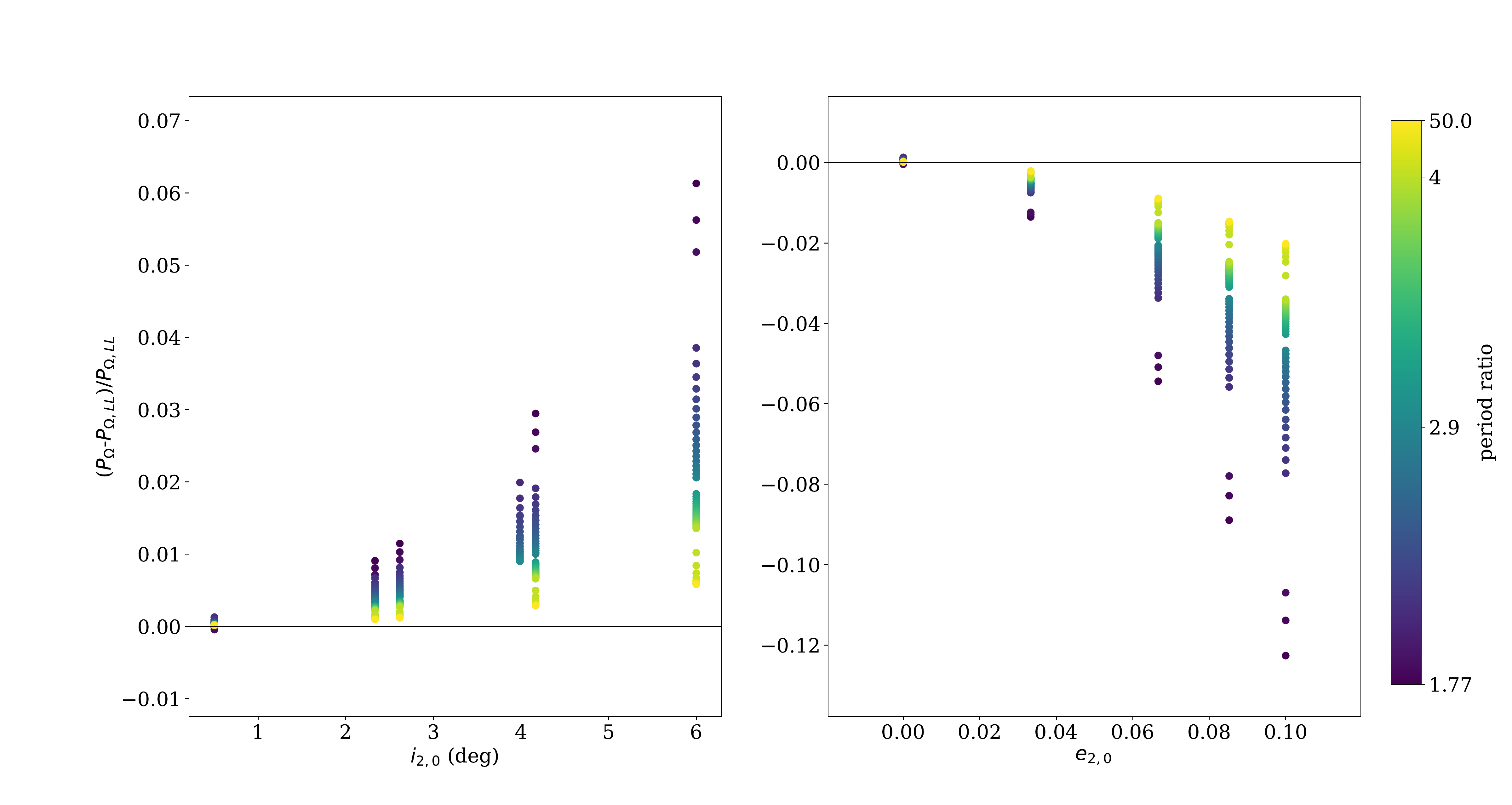}
  \caption{Fractional difference of simulated nodal precession period from linear theory ($P_{\text{LL}}=\frac{2\pi}{|\dot{\Omega}_{\text{LL}}|}$) of a test particle external to a massive planet as a function of (left panel:) inclination at zero eccentricity, and (right panel:) eccentricity at low inclination (0.5$^{\circ}$). Period ratios are denoted with colors with a break in scale at $P_2/P_1=4$.}
  \label{fig:simse0i0}
\end{figure*}

These results are shown in the two limiting cases (circular and the lowest inclination of 0.5$^\circ$) in Figure~\ref{fig:simse0i0}. Note the color bar has a break in scale at $P_2/P_1=4$ to better show the range of period ratios, which was more densely sampled for $P_2/P_1<4$. Zero on the vertical axis corresponds to the predicted value according to Laplace-Lagrange theory. The full set of results can be found in Table~\ref{apptab:restricted_results}.

The parameter space in period ratio was chosen to avoid first- and second-order mean motion resonances, where the secular interactions can no longer be considered as averaged over the orbits without considering their phase correlation. There was one third-order mean motion resonance in the period ratio grid at $P_2/P_1=4$. These simulations exhibited similar qualitative behavior but with a weaker $\alpha$ dependence than the other period ratio simulations, indicating that even third-order mean motion resonances can dominate over secular effects. The $P_2/P_1=4$ simulations were excluded from all future analyses and are not plotted in any figures.

\subsection{Unrestricted Three Body Simulations}

Given the marked deviations seen in the restricted simulations, we expanded our numerical approach to full three body simulations with two massive planets.

\subsubsection{Initial Conditions, Integration, and Analysis}\label{sec:ICintanalysis_U3b}

As before, initial system properties were randomly generated and then used for all simulations. The stellar mass was chosen uniformly between 0.3 and 1 $M_\odot$. The planets' masses were chosen uniformly between 2 and 10 $M_\oplus$. The inner planet's period was chosen uniformly between 4 and 8 days. The arguments of pericenter were chosen uniformly between -$\pi$ and $\pi$.

The same values of period ratio, eccentricity, and inclination were used as described in Section~\ref{sec:restsims}, where here both planets are given the same eccentricity $e$ and $i$ is the mutual inclination. Before simulating, the system was rotated into the invariable plane such that the total angular momentum vector is in the $z$ direction (neglecting spin angular momentum, as each body is modeled as a point mass).

\begin{table}[h!]
\centering
\caption{Initial conditions for the three body simulations with two massive planets.}
\label{tab:simICs_3body}
\begin{tabular}{|l|l|l|}
\hline
Stellar Mass & \multicolumn{2}{l|}{0.788 $M_\odot$} \\ \hline
\textbf{Planet:} & \textbf{b} & \textbf{c} \\ \hline
Period (days) & 6.2448 & ($P_2/P_1$)$\times$6.2448 \\ \hline
Mass ($M_\oplus$) & 7.6086 & 8.3235 \\ \hline
Eccentricity & $e$ & $e$ \\ \hline
$\omega$ (rad) & 2.5540 & 2.1423 \\ \hline
$\Omega$ (rad) & $\pi$ & 0 \\ \hline
Mutual Inclination & \multicolumn{2}{l|}{$i$} \\ \hline
\textbf{Variable Parameters} & \textbf{Min} & \textbf{Max} \\ \hline
$P_2/P_1$ & 1.77 & 50 \\ \hline
$e$ & 0 & 0.1 \\ \hline
$i$ (deg) & 0.5 & 6 \\ \hline
\end{tabular}
\end{table}

The initial conditions are shown in Table~\ref{tab:simICs_3body}. As with the restricted simulations, the values of $\Omega_{1,2}$ were saved every 0.4$P_1$, and each system was integrated for 10$^5P_2$ using \texttt{REBOUND} with IAS15 integrator. 

The nodal precession period was analyzed as described in Section~\ref{sec:simanal} for both the inner planet and the outer planet.

Due to some time-dependent effects (see Section~\ref{sec:timedepend}), inaccuracies arose in the higher period ratio systems because only a partial period of nodal precession was simulated in the 10$^5P_2$ time. In order to simulate a long enough time period to observe multiple cycles of nodal precession, the simulations with $P_2/P_1>$5 were extended to five times the initially-fit nodal period. To reduce computational time and memory, these longer simulations were done using the WHFast integrator \citep{2015Rein2} with timestep = 0.02$P_1$ and saving outputs approximately every 16000 steps. Because this integrator can introduce additional error in secular frequencies \citep{2019Rein}, we ran a comparison with the previous IAS15 results. This comparison showed that the use of WHFast and decreased sampling had negligible effect on the accuracy of the nodal precession period, agreeing within 10$^{-5}$.

\subsubsection{Results}\label{sec:numres_3body}

\begin{figure*}[h!]
  \includegraphics[width=\linewidth]{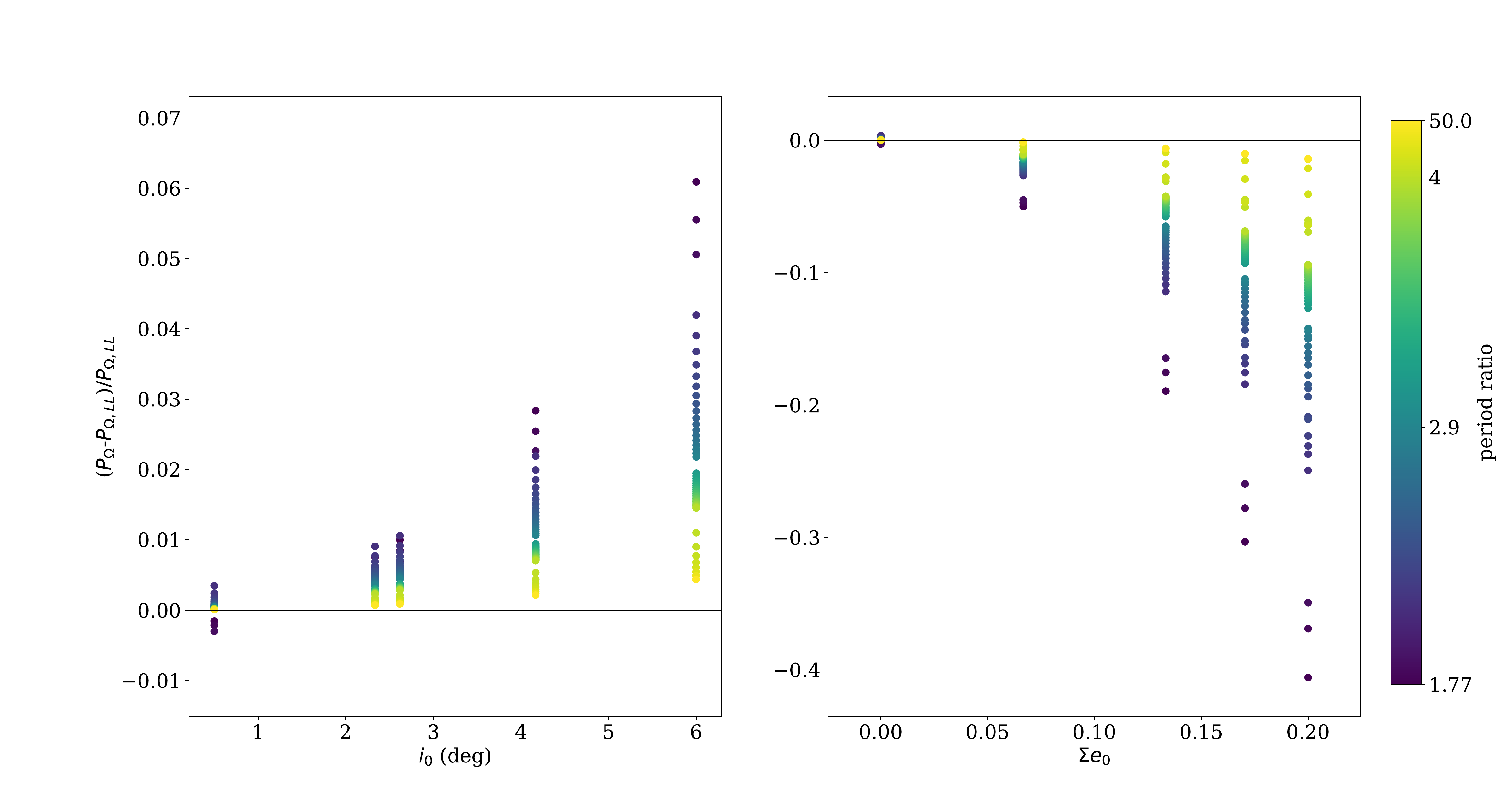}
  \caption{Fractional difference of simulated nodal precession period from linear theory in a system with two massive planets as a function of (left panel:) mutual inclination at zero eccentricity, and (right panel:) combined eccentricity ($\Sigma e_{0}=e_{1,0}+e_{2,0}=2e$; see Table~\ref{tab:simICs_3body}) at low mutual inclination (0.5$^{\circ}$). Period ratios are denoted with colors with a break in scale at $P_2/P_1=4$. Plotted are the results for the outer planet; the inner planet data is similar.}
  \label{fig:simse0i0_3body}
\end{figure*}

The fractional deviation from the linear theory period is plotted in Figure~\ref{fig:simse0i0_3body}. Only the outer planet is plotted in each panel, but the inner planet data is similar, as expected. The full set of results can be found in Table~\ref{apptab:unrestricted_results}.

The nodal precession periods of the unrestricted three body simulations are similar to the restricted simulations in Section~\ref{sec:numres_rest}. As $\alpha$ increases (period ratio decreases), the eccentricity and inclination dependence also increases. 

There is a greater range in the effect of eccentricity here than compared to the test particle case. This is expected because here both the inner and outer planet have eccentricity, whereas in the restricted case only the test particle was given eccentricity. To better illustrate this difference and for comparison with the restricted case, we have used combined eccentricity, $\Sigma e_{0}=e_{1,0}+e_{2,0}$, to plot results for the unrestricted systems. Our simulations use equal eccentricities for the unrestricted planets, and therefore do not examine the effect of the partitioning of the total eccentricity between both planets. The scale of the effect is similar when considering only combined eccentricity in the same range as the test particle's eccentricity ($<$0.10).

\section{Analytical Investigation}\label{sec:analytical}

As is clearly seen in the results of our numerical simulations (Sections \ref{sec:numres_rest}, \ref{sec:numres_3body}), including only low order terms of $e$, $i$, and $\alpha$ is not adequate to describe the nodal precession period of our simulated systems. Using \cite{MurrayDermott}, we developed a higher order expression for $\dot{\Omega}$. 

\cite{MurrayDermott} Equation 6.148 shows the change in the ascending node over time. This equation can be combined with the disturbing function in their Equation 6.44/6.45 and the fourth-order expansion in $e$ and $i$ of the disturbing function parts in their Appendix B, solved for secular case of $j=0$ and averaged over the orbits (i.e., neglecting terms containing mean longitudes in the cosine arguments). The resulting $\dot{\Omega}$ expressions are shown in Equations \ref{eqn:Omdot1} and \ref{eqn:Omdot2} with the disturbing function derivatives shown in Equations \ref{eqn:dRDds1} and \ref{eqn:dRDds2}.

\begin{equation}\label{eqn:Omdot1}
    \dot{\Omega}_1 = \frac{m_2 n_1 \alpha}{4 (m_1+M_\star) s_1 \sqrt{1-e_1^2}} \frac{\partial R_D}{\partial s_1}
\end{equation}

\begin{equation}\label{eqn:Omdot2}
    \dot{\Omega}_2 = \frac{m_1 n_2}{4 (m_2+M_\star) s_2 \sqrt{1-e_2^2}} \frac{\partial R_D}{\partial s_2}
\end{equation}

\begin{multline}\label{eqn:dRDds1}
    \frac{\partial R_D}{\partial s_1} = 2 s_1 f_3 + 2 s_1 (e_1^2+e_2^2) f_7 + 4 s_1^3 f_8 + 2 s_1 s_2^2 f_9 - (2 s_1 f_{13} - s_2 f_{22} - s_2 f_{23}) e_1 e_2 cos(\omega_2-\omega_1)\\ - s_2 f_{14} - (e_1^2+e_2^2) s_2 f_{15} - (3 s_1^2 s_2 + s_1^3) f_{16} + (2 s_1 f_{18} - s_2 f_{21}) e_1^2 cos(2 \omega_1)\\ - (2 s_1 f_{19} - s_2 f_{24}) e_1 e_2 cos(\omega_2 + \omega_1) + (2 s_1 f_{20} - s_2 f_{25}) e_2^2 cos(2\omega_2) + 2 s_1 s_2^2 f_{26}
\end{multline}

\begin{multline}\label{eqn:dRDds2}
    \frac{\partial R_D}{\partial s_2} = 2 s_2 f_3 + 2 s_2 (e_1^2+e_2^2) f_7 + 4 s_2^3 f_8 + 2 s_2 s_1^2 f_9 - (2 s_2 f_{13} - s_1 f_{22} - s_1 f_{23}) e_1 e_2 cos(\omega_2-\omega_1)\\ - s_1 f_{14} - (e_1^2+e_2^2) s_1 f_{15} - (3 s_2^2 s_1 + s_2^3) f_{16} + (2 s_2 f_{18} - s_1 f_{21}) e_1^2 cos(2 \omega_1)\\ - (2 s_2 f_{19} - s_1 f_{24}) e_1 e_2 cos(\omega_2 + \omega_1) + (2 s_2 f_{20} - s_1 f_{25}) e_2^2 cos(2\omega_2) + 2 s_2 s_1^2 f_{26}
\end{multline}

Here, $s_{1,2} = $sin$(i_{1,2}/2)$. The $f$ equations are functions of $\alpha$ only and are given in Appendix~\ref{apsec:ffuncs}.

Because of the $s_{1,2}$ derivative and denominator term in Equations~\ref{eqn:Omdot1} and \ref{eqn:Omdot2}, the highest order of $s_{1,2}$ is 2 in the expression for $\dot{\Omega}_{1,2}$. Similarly, the highest order of $e_{1,2}$ is 2. The highest explicit order of $\alpha$ is 5, with additional $\alpha$ dependence arising from the Laplace coefficients (see Appendix~\ref{apsec:ffuncs}).

For a 2-planet system, $\dot{\Omega}_1$ and $\dot{\Omega}_2$ are expected to be equal. Using this analytical solution, $\dot{\Omega}_1 \simeq \dot{\Omega}_2$ within about 0.1\% for our simulations, a good check on the derivation. Additionally, this means the use of restricted test particle simulations, where only $\dot{\Omega}_2$ is calculated, provides reasonably generalizable results.

\subsection{Analytical Results}\label{sec:analyticalres}

\begin{figure*}[h]
  \includegraphics[width=\linewidth]{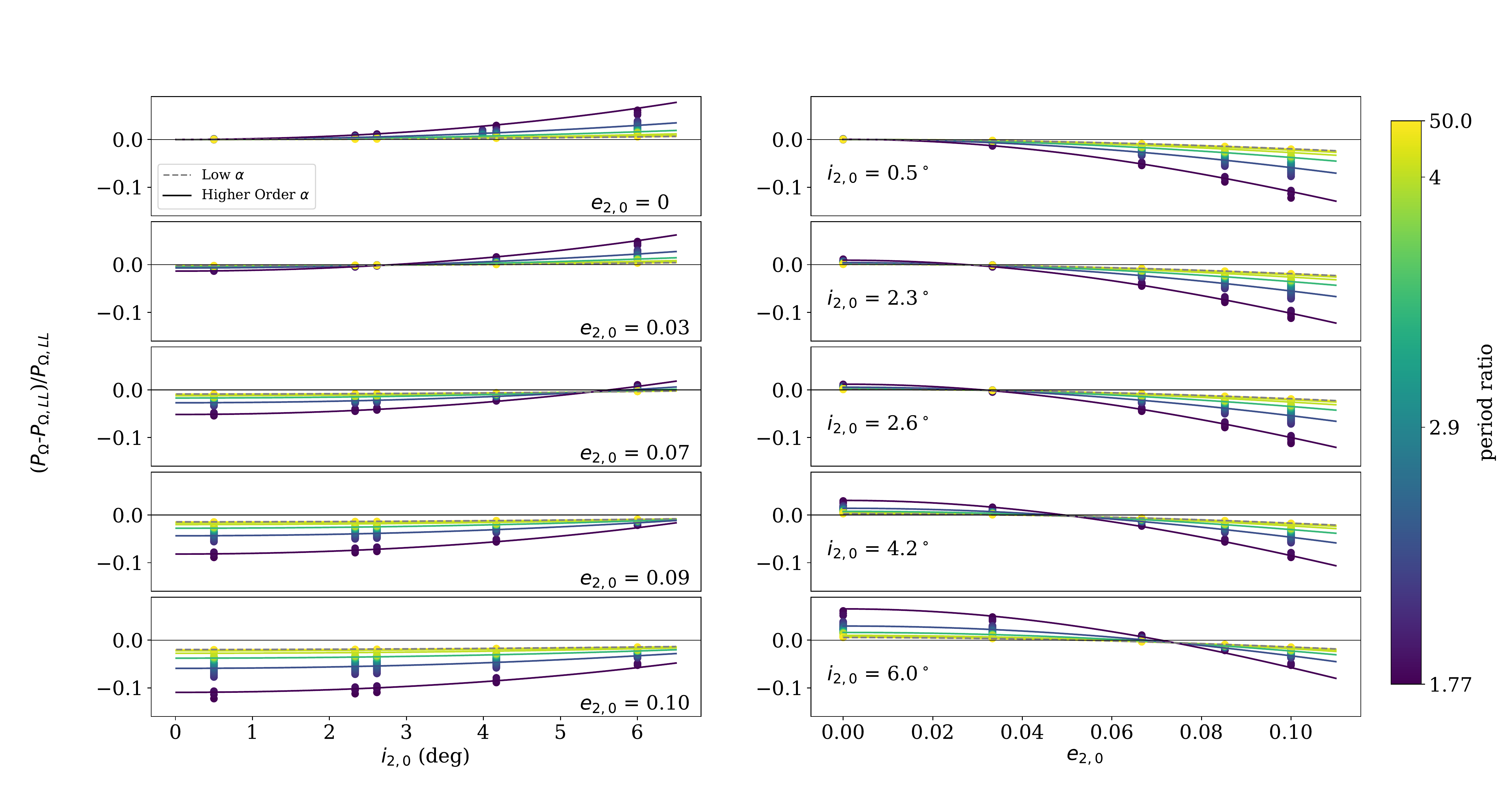}
  \caption{Fractional difference of simulated nodal precession period from linear theory of a test particle external to a massive planet as a function of (left panel:) inclination at various eccentricities, and (right panel:) eccentricity at various inclinations. Period ratios are denoted with colors with a break in scale at $P_2/P_1=4$. Points show the data from simulations while the solid lines show the results from the analytical solution from Equation~\ref{eqn:Omdot2} ($P_\Omega=\frac{2\pi}{|\dot{\Omega}|}$) at a sampling of different period ratios. The dashed gray line shows the analytical solution from Equation~\ref{eqn:LW11eqn33}.}
  \label{fig:data_MD4_all}
\end{figure*}

Calculating the expected nodal precession period with the fourth-order derivation and comparing it to our restricted test particle simulations from Section~\ref{sec:restsims}, we find that this approximation is much improved from Equation~\ref{eqn:LW11eqn33}. Figure~\ref{fig:data_MD4_all} shows a sampling of calculated $P_\Omega$ for various values of $\alpha$ along with the data from the restricted simulations. This figure is similar to Figure~\ref{fig:simse0i0}, but showing a wider range of combinations of inclination and eccentricity as well as the calculated fourth-order model and the second-order prediction from Equation~\ref{eqn:LW11eqn33}.

\begin{figure*}[h]
  \includegraphics[width=\linewidth]{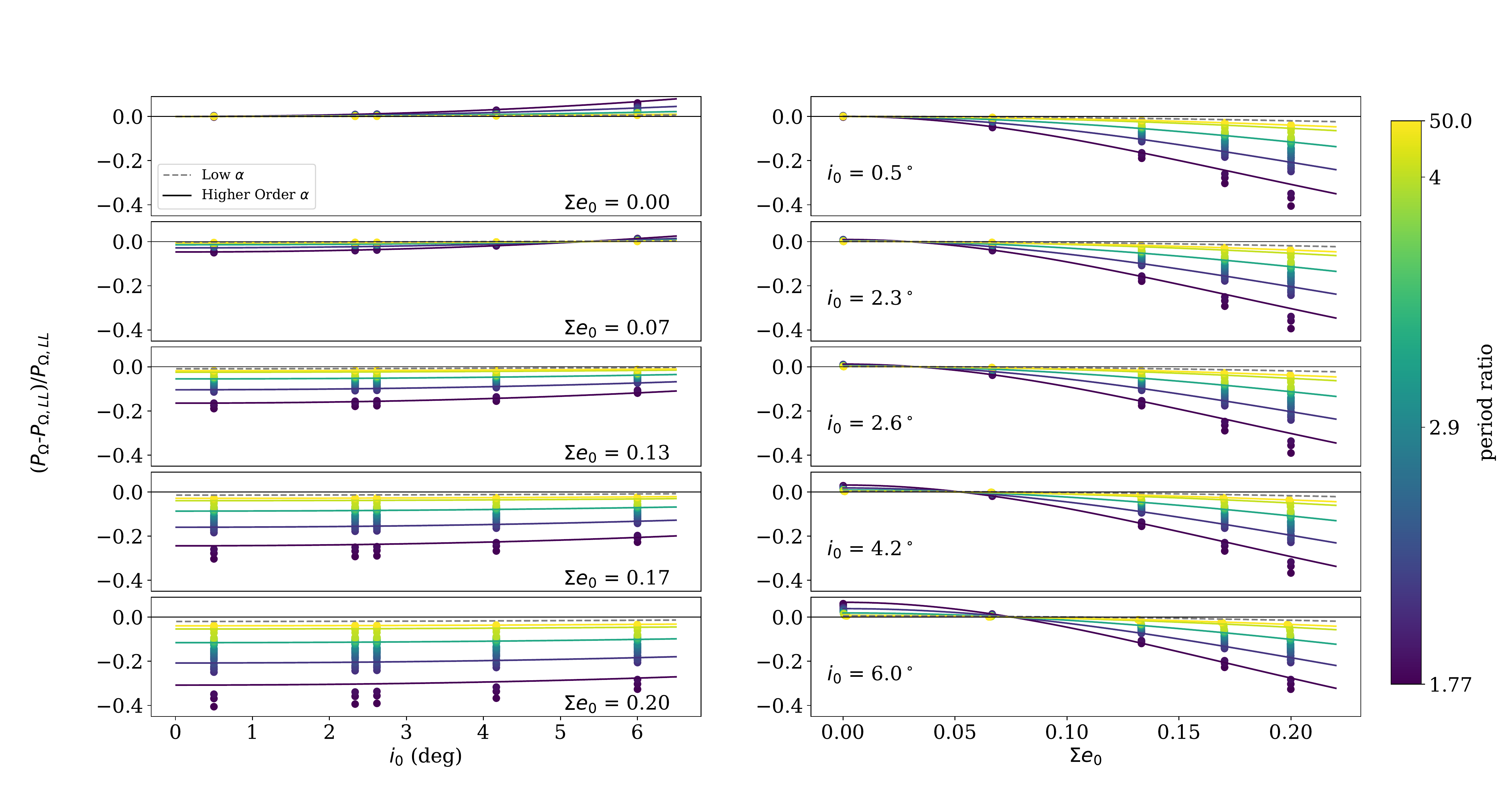}
  \caption{Nodal precession rate variation in a system with two massive planets as a function of (left panel:) mutual inclination at various combined eccentricities, and (right panel:) combined eccentricity at various mutual inclinations. Period ratios are denoted with colors with a break in scale at $P_2/P_1=4$. Plotted are the results for the outer planet, the inner planet data is similar. Points show the data from simulations while the solid lines show the results from the analytical solution from Equation~\ref{eqn:Omdot2} at a sampling of different period ratios. The dashed gray line shows the analytical solution from Equation~\ref{eqn:LW11eqn33}.}
  \label{fig:data_MD4_all_3body}
\end{figure*}

The analytical solution approaches that of \cite{2011Lithwick} for low $\alpha$ and matches the stronger inclination and eccentricity dependence that we see for high $\alpha$. However, because this is a restricted test particle case, the calculation is greatly simplified by the fact that $m_2=e_1=s_1=0$ and only $\dot{\Omega}_2$ is calculated. To better evaluate the accuracy of the full analytical solution, we evaluate the comparison with the full unrestricted three body simulations. The results are shown in Figure~\ref{fig:data_MD4_all_3body}.

Here we see that the analytical solution is again a reasonable descriptor of the simulation behavior (with a median accuracy of 0.11\%), although the highest-$\alpha$ systems deviate more strongly than predicted, presumably due to higher-order effects. The low-$\alpha$ solution still approximates Equation~\ref{eqn:LW11eqn33}, although not as exactly as in the test particle case. This difference is not unexpected as \cite{2011Lithwick} consider a test particle and a circular massive perturber, whereas in this case both planets have mass and eccentricity; we use mutual inclination and average eccentricity in calculating Equation~\ref{eqn:LW11eqn33} to account for this change. The results for the inner and outer planets are almost functionally identical, as expected. Only the outer planet results are plotted in Figure~\ref{fig:data_MD4_all_3body} for clarity.

\begin{figure}[H]
  \floatbox[{\capbeside\thisfloatsetup{capbesideposition={left,center}}}]{figure}[1.2\FBwidth]
  {\caption{Calculated values for the Laplace coefficients terms with the lowest order of $\alpha$ that appears in Equations~\ref{eqn:Omdot1} to \ref{eqn:dRDds2}. The black line shows the $b_{3/2}^{(1)}\alpha$ term that appears in both this solution and in \cite{2011Lithwick} as the $3\alpha^2$ approximation (black dashed line). The red dotted lines show the $b_{5/2}^{(j)}\alpha^2$ terms, with the darker shades showing lower values of $j$. The blue dashed-dotted lines show the same for the $b_{7/2}^{(j)}\alpha^3$ terms.}}
  {\includegraphics[width=\linewidth]{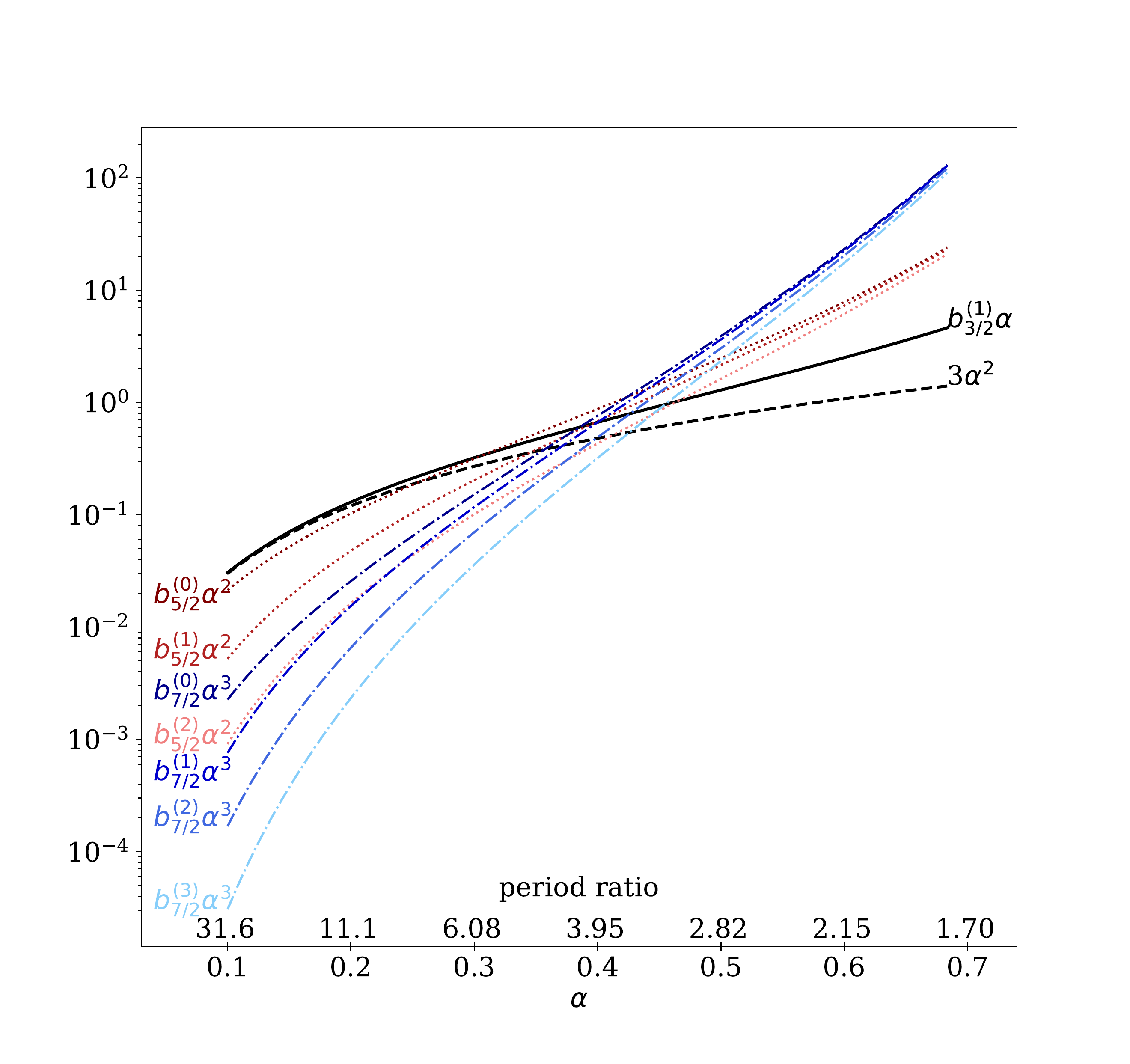}}
  \label{fig:LPC_alpha}
\end{figure}

A direct comparison of this derivation with that of \cite{2011Lithwick} requires combining their Equations 1, 6, 15, and 33 to obtain an expression for $\dot{\Omega}$ and comparing that to this paper's Equations \ref{eqn:Omdot1} and \ref{eqn:dRDds1} with $m_1=0$, $s_2=0$, and $e_2=0$. The general form of the solutions is similar, indicating dependence on $-e_1^2$ and $s_1^2$. The primary difference arises from the higher-order $\alpha$ terms in the $f$ functions (see Appendix \ref{apsec:ffuncs}) and the explicit calculation of the Laplace coefficients. Figure~\ref{fig:LPC_alpha} shows how terms that are negligible at low $\alpha$ become much stronger at higher $\alpha$, as well as how the $b_{3/2}^{(1)}=\alpha$ approximation loses accuracy. Another difference arises from the dependence on $\omega_1$, but this is negligible when $e_2=0$ (see Section \ref{sec:timedepend} for details). Additionally, the \cite{2011Lithwick} solution is only valid for a test particle and a circular massive planet, and several additional terms arise when considering the more general case of two planets with inclinations and eccentricities, including a significant dependence on the relative orientation of the orbits $(\omega_2-\omega_1)$.

It should be noted that Equations~\ref{eqn:Omdot1} to \ref{eqn:dRDds2} use orbital elements taken relative to the central star (astrocentric). The default orbital elements in \texttt{REBOUND} are Jacobian. However, comparing the calculated analytical results for using Jacobian versus astrocentric coordinates in the unrestricted system has an effect no larger than 0.012\% and on average 0.015\%, so we did not recalculate the previously saved Jacobian elements and used them for all calculations. This effect could be more significant in other systems, particularly those with a higher planet-to-star mass ratio, and it is recommended to use astrocentric elements when applying the analytical solution.

\subsection{Time Dependence}\label{sec:timedepend}

The expressions for $\dot{\Omega}_1$ and $\dot{\Omega}_2$ depend on terms that change over time. In the secular regime, the semi-major axes (and therefore $\alpha$) are constant, but inclination, eccentricity, and $\omega$ vary periodically. The change in inclination is small, but the variations in eccentricity and $\omega$ can be significant. The result of this change in orbital elements is to change the rate of nodal precession over time.

Using data from the three body simulations over time, we were able to examine the magnitude of this change by calculating the instantaneous analytical solution at each timestep. The data showed that the term contributing the most time variation arises from $\omega_2+\omega_1$, and that the mean over time is driven by $\omega_2-\omega_1$. Because the time variations are periodic on a shorter timescale than the nodal precession period, the mean analytical solution over time is the observed nodal precession period. Using the orbital elements from a single point in time can then produce an instantaneous analytical solution that varies from the mean underlying period by a potentially significant amount. However, this error is easily eliminated by removing the terms that vary on short timescales (these are the terms with $\text{cos}(2\omega_1)$, $\text{cos}(\omega_2+\omega_1)$, and $\text{cos}(2\omega_2)$ in Equations~\ref{eqn:Omdot1}/\ref{eqn:Omdot2}). This method of calculating $\dot{\bar{\Omega}}$ has been used to calculate all the higher-order-$\alpha$ analytical solutions used in this paper.

\begin{figure}[H]
  \floatbox[{\capbeside\thisfloatsetup{capbesideposition={right,center}}}]{figure}[1.2\FBwidth]
  {\includegraphics[width=\linewidth]{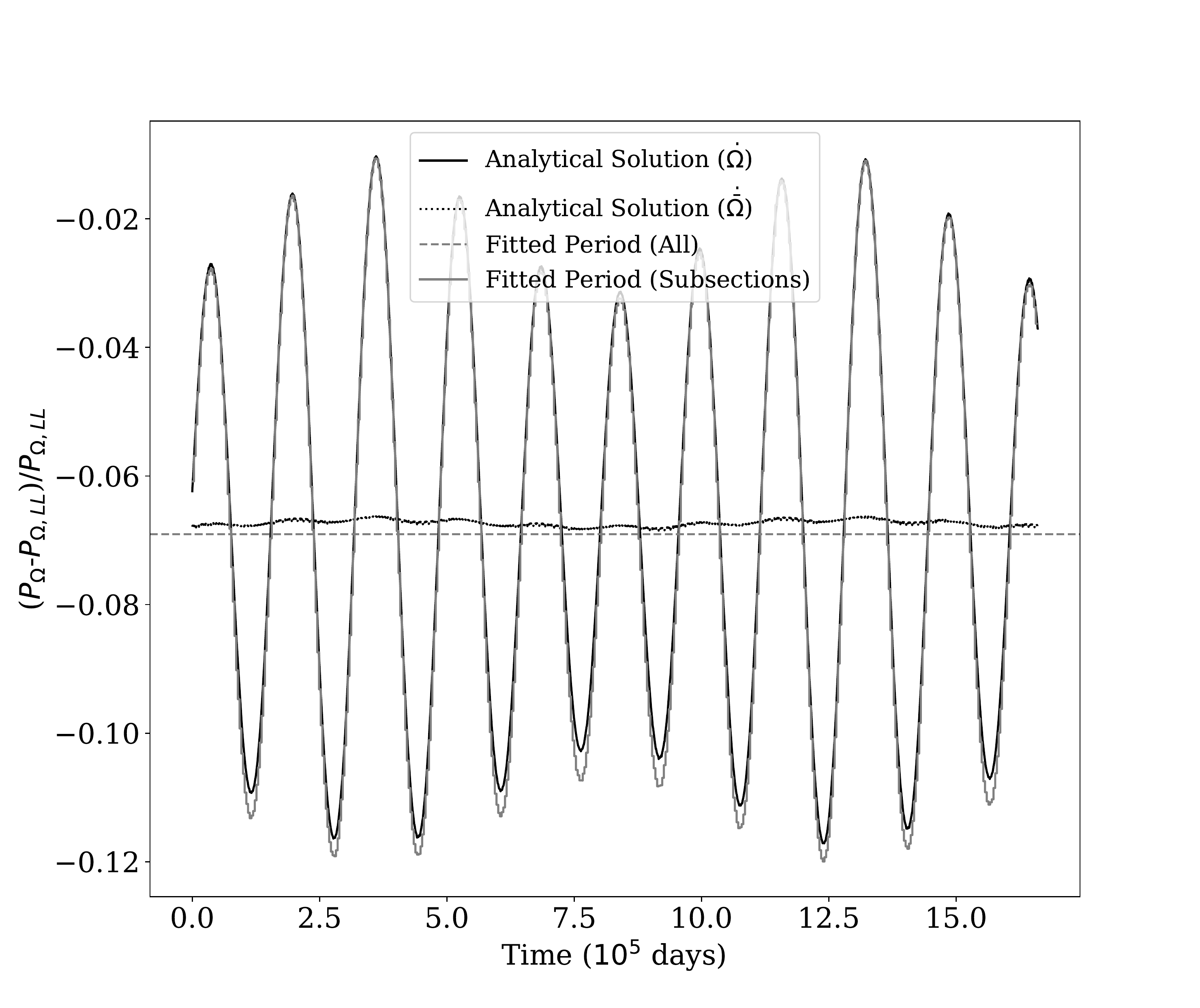}}
  {\caption{An example of how the analytical and simulated precession rate changes over time. For this simulation, $e_{0}=e_{1,0}=e_{2,0}=$0.07, $i_0$=2.6$^\circ$, and $P_2/P_1=2.6590$. The instantaneous fourth-order analytical solution at each timestep is shown in solid black. The mean fourth-order analytical solution at each timestep is shown in dotted black. The gray dashed line shows the period fitted from the analysis in Section~\ref{sec:simanal} (this is the value plotted in Figure~\ref{fig:data_MD4_all_3body}). The solid gray lines show a value for a fitted period within a shorter subset of the simulation time. The time variations are on a shorter timescale than the nodal precession period, $P_\Omega\approx5.7\times10^5$ days.}}
  \label{fig:period_overtime}
\end{figure}

The variation of $\dot{\Omega}$ is also seen in the simulated data, confirmed by fitting periods to subsets of $\Omega$ over time. The period found via the analysis in Section~\ref{sec:simanal} by fitting to the entire output dataset is the average period. As expected, this average period is very well-defined given the difference in time scales. An example from one simulation is shown in Figure~\ref{fig:period_overtime} to illustrate. Using the mean analytical solution as described greatly reduces the time-variation of the calculated solution and provides an accurate average period from the system parameters at a single point in time.

\subsection{Effect of General Relativity}

General relativity predicts that the curvature of spacetime can lead to apsidal precession of planetary orbits \citep{1916Einstein}. Given the dependence of the $\dot{\Omega}$ on $\omega$, it is possible for this effect to couple to the nodal precession rate.

Calculating the predicted apsidal precession from general relativity and comparing it to the simulated apsidal precession rate, the general relativity effect only becomes comparable for period ratios $\gtrsim$10. Given that the effect of $\omega$ is much reduced at high period ratios, due to the strong $\alpha$ dependence in those terms, neglecting general relativity (or other sources of extra apsidal precession) in our simulations is not expected to have a significant effect on our results.

\section{Limits of Applicability}\label{sec:applicablerange}

In many cases, the simpler nodal precession period given by Laplace-Lagrange theory is adequate. However, for cases where a more precise period is desired, a higher order approximation might be needed. To find the parameter space where the higher-order-$\alpha$ analytical solution given in Section~\ref{sec:analytical} attains a high accuracy, over 9900 additional systems with higher inclinations and eccentricities were simulated. These systems were the same as those described in Section~\ref{sec:ICintanalysis_U3b}, except with a wider range of eccentricities and inclinations, going as high as 1.8 in combined eccentricity and 30 degrees in mutual inclination, although the majority were at combined eccentricities of $\leq$0.4 to determine the location of the transition for analytical accuracy. Many of the $\Sigma e_0 > 0.4$ systems were not stable, changing rapidly in semi-major axis.

\begin{figure*}[h]
  \includegraphics[width=\linewidth]{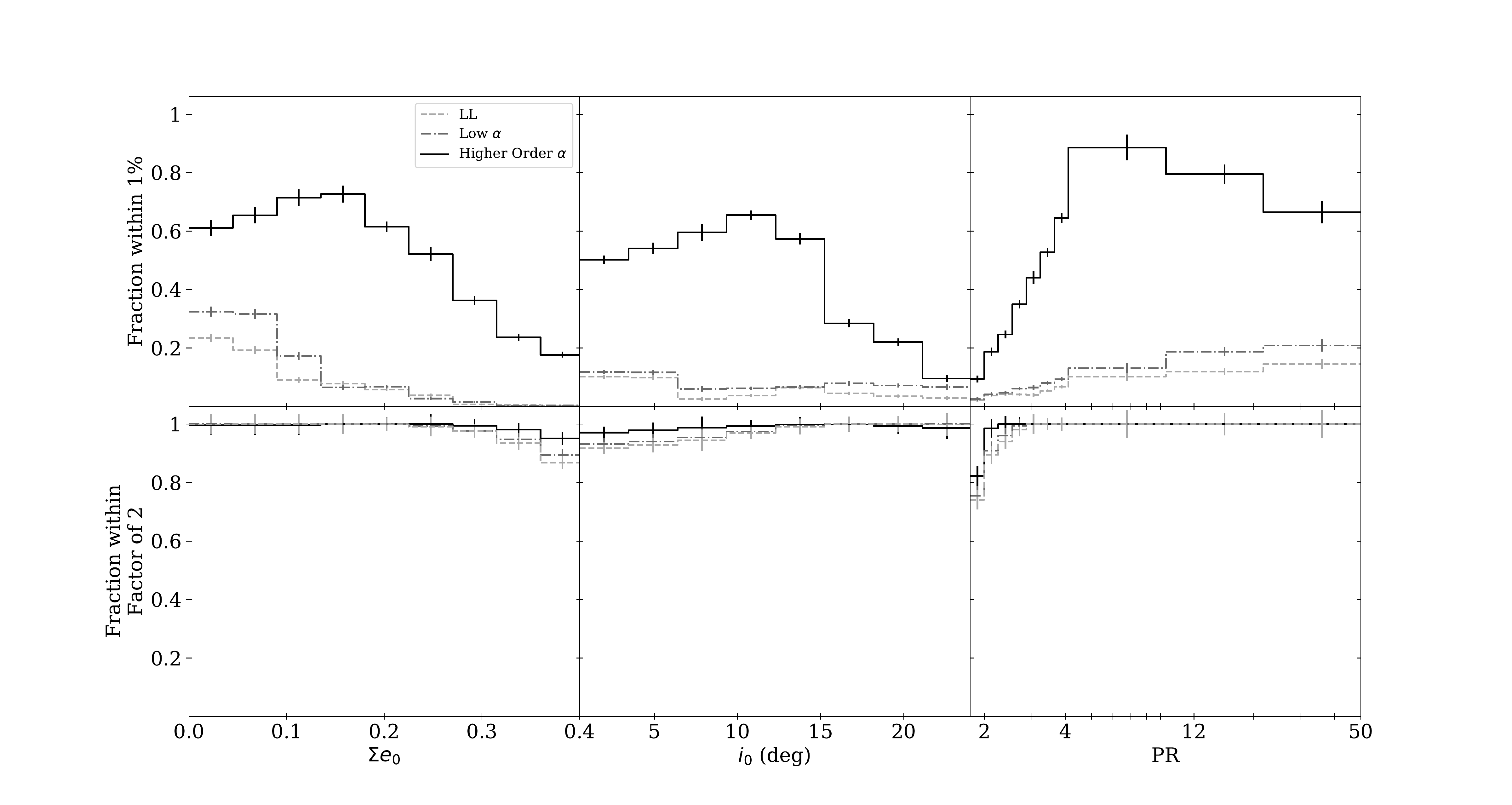}
  \caption{The fraction of simulations per bin that are below a given threshold of accuracy, where accuracy is calculated as $(P_{\Omega,sim}-P_{\Omega,model})/P_{\Omega,model}$ for three models: Laplace-Lagrange theory, low-$\alpha$ approximation \citep{2011Lithwick}, higher-order-$\alpha$ approximation (Section~\ref{sec:analytical}). Error bars on each bin are calculated from a binomial distribution.}
  \label{fig:iae_accuracy}
\end{figure*}

Figure~\ref{fig:iae_accuracy} shows the fraction of simulations per bin in combined eccentricity, mutual inclination, and period ratio space that were under a given threshold of accuracy. These distributions are for only for our one particular simulated system (see Section~\ref{sec:ICintanalysis_U3b}), and a simple one-dimensional distribution does not capture the more complicated interplay between the three parameters. However, this plot does provide a general guideline for when certain approximations are sufficient. As expected, the accuracy decreases for larger eccentricity and inclination and for smaller period ratio (larger $\alpha$).

When high accuracy, within 1\%, is desired, the higher-order-$\alpha$ approximation (Equations~\ref{eqn:Omdot1}/\ref{eqn:Omdot2}) is needed, especially for low period ratios ($\lesssim$4). For very low period ratios ($\lesssim$2.1), an N-body simulation is likely required unless the eccentricities and inclinations are low. The higher-order-$\alpha$ approximation is typically accurate for mutual inclinations $\lesssim$15$^\circ$ and for a combined eccentricity $\lesssim$0.27. Across all stable simulations, the median accuracy of the higher-order-$\alpha$ approximation is 1.37\%.

The choice of 1\% as the threshold for high accuracy can be compared with scales of interest for application. For example, imposing a 1\% error on the nodal precession frequency values when calculating the secular resonance location in the \object[2MASS J19151032+4802248]{\emph{Kepler}-117} system (see Section~\ref{sec:secularres} for details) produces the same order of magnitude error that arises from observational uncertainty.

The location of the secular resonance, even with exactly known input parameters, has an intrinsic width. This width gives another scale for the accuracy of the nodal precession frequency, and here we have calculated it by finding where the inclination of a test particle is forced to three times its natural inclination (i.e., its forced inclination if the two perturbing planets were on fixed orbits) using linear theory. The width varies based on the system configuration. For example, in the {\emph{Kepler}-117} system, the secular resonance widths are of order 0.1 times the resonance location. A random 1\% error on the nodal precession frequency would then always place the calculated resonance location inside of the actual resonance, with inaccurate placement beginning at about 3\% error on the nodal precession frequency. To check the generality of this, we generated 100 random 2-planet systems. The relative resonance width (the width of the resonance divided by its location) was typically 0.1-0.2, with the inner resonance usually being larger by a factor of about 2-3, similar to the {\emph{Kepler}-117} system. High-$\alpha$ systems tended to the lower end of the range. Thus a 1\% accuracy on the nodal precession frequency is adequate to ensure correctly identifying the predicted location of the secular resonance.

\begin{figure*}[h]
\centering
  \includegraphics[width=\linewidth]{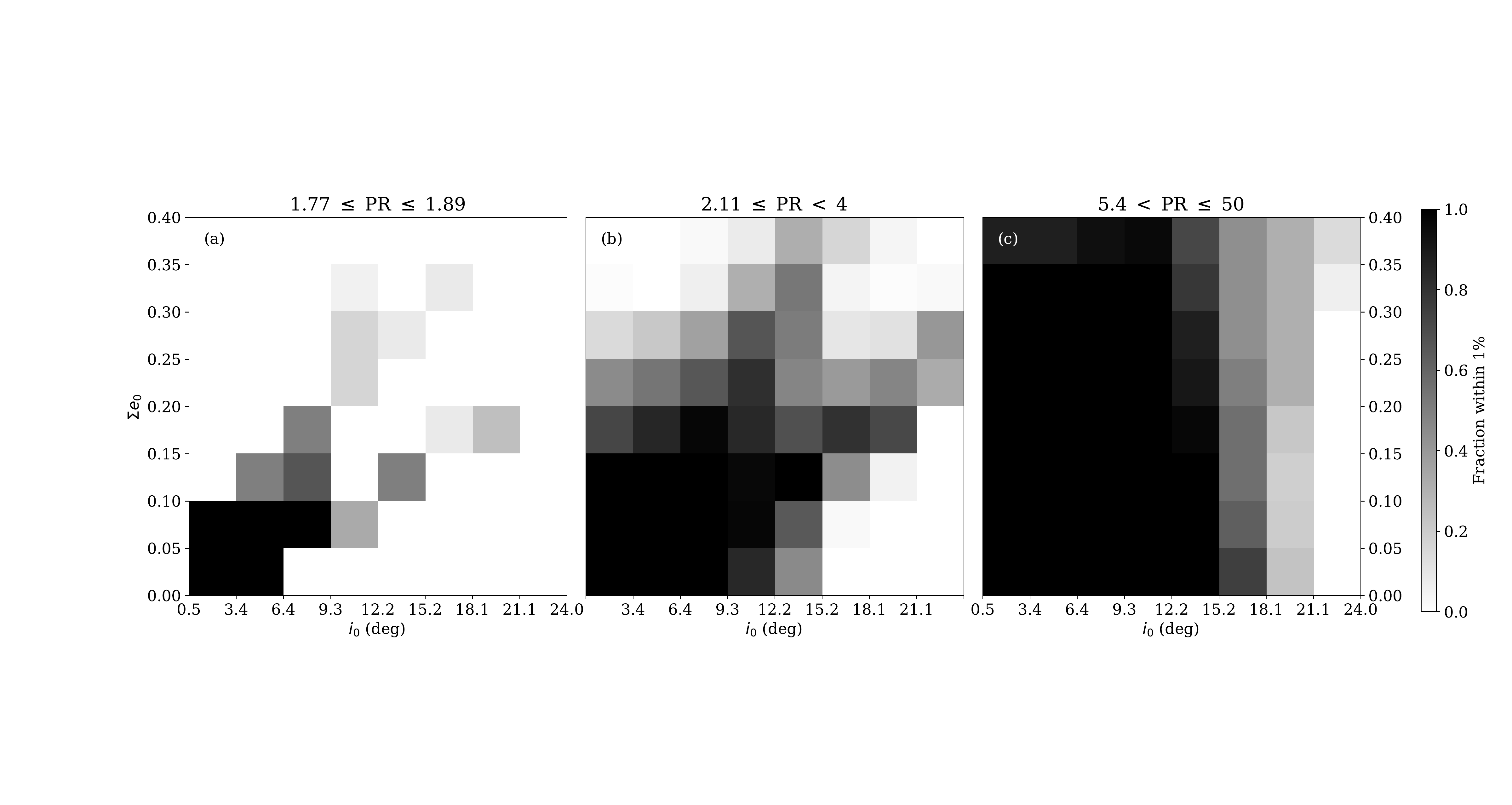}
  \caption{The fraction of simulations per bin that are accurate to within 1\%, where accuracy is calculated as $(P_{\Omega,sim}-P_{\Omega,model})/P_{\Omega,model}$ for the higher-order-$\alpha$ approximation (Section~\ref{sec:analytical}). Shown for three different ranges of period ratio: (a) 1.77 to 1.89, (b) 2.11 to 4, and (c) 5.4 to 50.}
  \label{fig:ei_hist2d}
\end{figure*}

A more detailed examination of the applicable parameter space is shown in Figure~\ref{fig:ei_hist2d}. For high period ratios, the fourth-order analytical solution is very accurate up to $\sim$18$^\circ$ mutual inclination and over 0.4 in combined eccentricity. For lower period ratios, the region of parameter space where the model is accurate shrinks to lower mutual inclination and combined eccentricity, as expected. There is also a region of moderate mutual inclination and combined eccentricity where the accuracy is very good. This region of accuracy arises due to the offsetting effects of inclination and eccentricity. For period ratios between 2.11 and 4, the model is accurate up to $\sim$15$^\circ$ mutual inclination and $\sim$0.2 in combined eccentricity. For very low period ratios, between 1.77 and 1.89, the model is accurate at mild mutual inclinations of $\leq6^\circ$ and combined eccentricities of 0.1.

If only an order of magnitude estimate is needed for the nodal precession period, then linear Laplace-Lagrange theory suffices over almost the entire range examined: mutual inclinations up to 30$^\circ$ and combined eccentricity up to 0.4. For higher eccentricities, the accuracy is typically limited not by the analytical approximation but by instability, and this region of parameter space was not well-sampled, particularly at higher period ratios. Across all stable simulations, the median accuracy of the linear theory solution is 12.8\%.

The consideration of nodal precession within a mean-motion resonance is beyond the scope of this paper. We have considered only period ratios well away from a mean-motion resonance, but have not carefully defined where the transition between resonant and non-resonant nodal precession behavior occurs. 

We have considered here primarily a single system with a low planet-to-star mass ratio ($\sim$6$\times 10^{-5}$) and planets of similar masses ($\sim$0.91 mass ratio). A small set of preliminary simulations of a system with the same planetary mass ratio but a higher planet-to-star mass ratio ($\sim$6$\times 10^{-3}$) and a system with the same planet-to-star mass ratio but differing planetary masses ($\sim$0.54 mass ratio) showed similar behavior to that seen here. Additionally, the dynamics of a planetary system are expected to be scale invariant, and so these effects are expected to be independent of the absolute semi-major axes of the planets. Thus we expect that these results are generally applicable.

\section{Application to \emph{Kepler} Population}\label{sec:Kepler}

\begin{figure*}[h]
  \includegraphics[width=\linewidth]{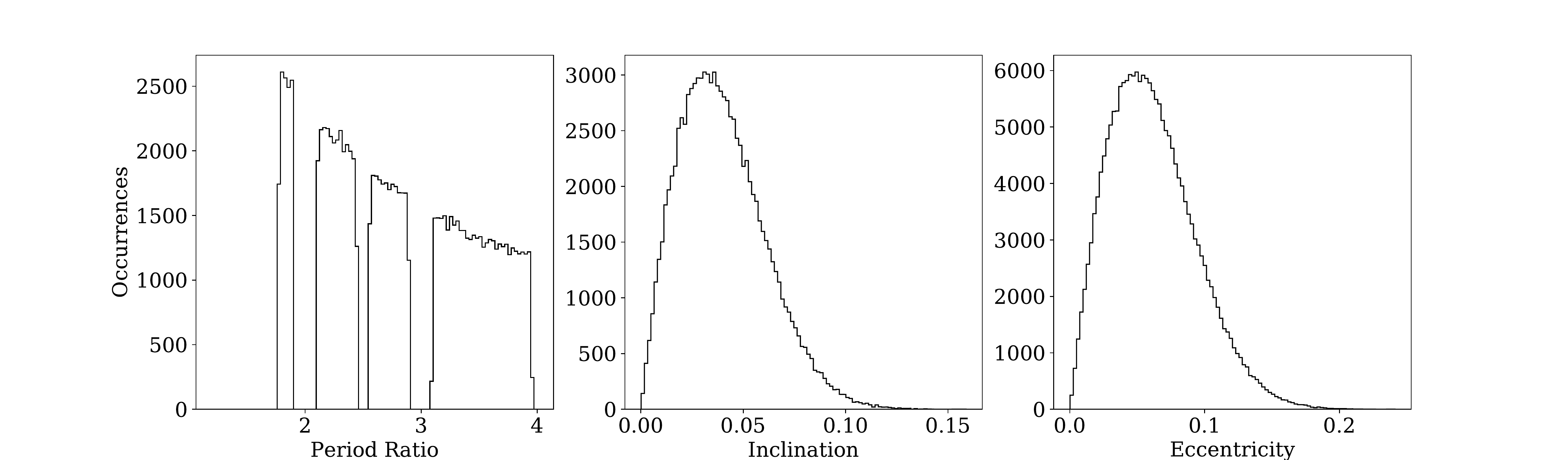}
  \caption{Distributions of period ratio, mutual inclination, and eccentricity drawn for a population of \emph{Kepler}-like multiplanet systems.}
  \label{fig:K3panel}
\end{figure*}

To estimate the effect on a population of multiplanet systems like those found in the \emph{Kepler} mission (e.g. \citealt{2014Fabrycky}), we drew 10$^5$ samples from a period ratio distribution, inclination distribution, and eccentricity distribution that describes the \emph{Kepler} multiplanet systems. The period ratio samples were drawn from a log-uniform distribution in between 1.2 and 4, resampling any period ratios that fell within 0.1 of a 1st or 2nd order resonance or within 0.05 of a 3rd order resonance (ensuring $|P_2/P_1-M/N| \geq 0.1$ for $M=N+1$ and $M=N+2$ and $|P_2/P_1-M/N| \geq 0.05$ for $M=N+3$). The mutual inclinations were drawn from a Rayleigh distribution with scale $\sigma=0.032$ radians \citep{2014Fabrycky}, and the eccentricities were drawn from a Rayleigh distribution with scale $\sigma=0.049$ \citep{2015VanEylen}. The resulting sample distributions are shown in Figure~\ref{fig:K3panel}.

In order to calculate the expected nodal precession deviation given by the analytical solution in Section~\ref{sec:analytical}, system properties including stellar and planetary masses, inner planet period, and arguments of pericenter are needed. For the purposes of finding trends from inclination, eccentricity, and semi-major axis ratio, the calculation was repeated 100 times for each set of $e$, $i$, and $\alpha$ with random draws of the remaining parameters. The effect of the masses and inner planet period are expect to scale away and should not affect the resulting calculation.

The period ratio was converted to $\alpha$ using Equation~\ref{eqn:alphaPR}. Each planet's eccentricity was drawn individually, and their inclinations were calculated from the mutual inclination such that the reference plane was the invariable plane. The analytical solution was calculated as described in Section~\ref{sec:timedepend} to remove the time-dependent effect of the arguments of pericenter.

\begin{figure}[H]
  \floatbox[{\capbeside\thisfloatsetup{capbesideposition={left,center}}}]{figure}[1.2\FBwidth]
  {\includegraphics[width=\linewidth]{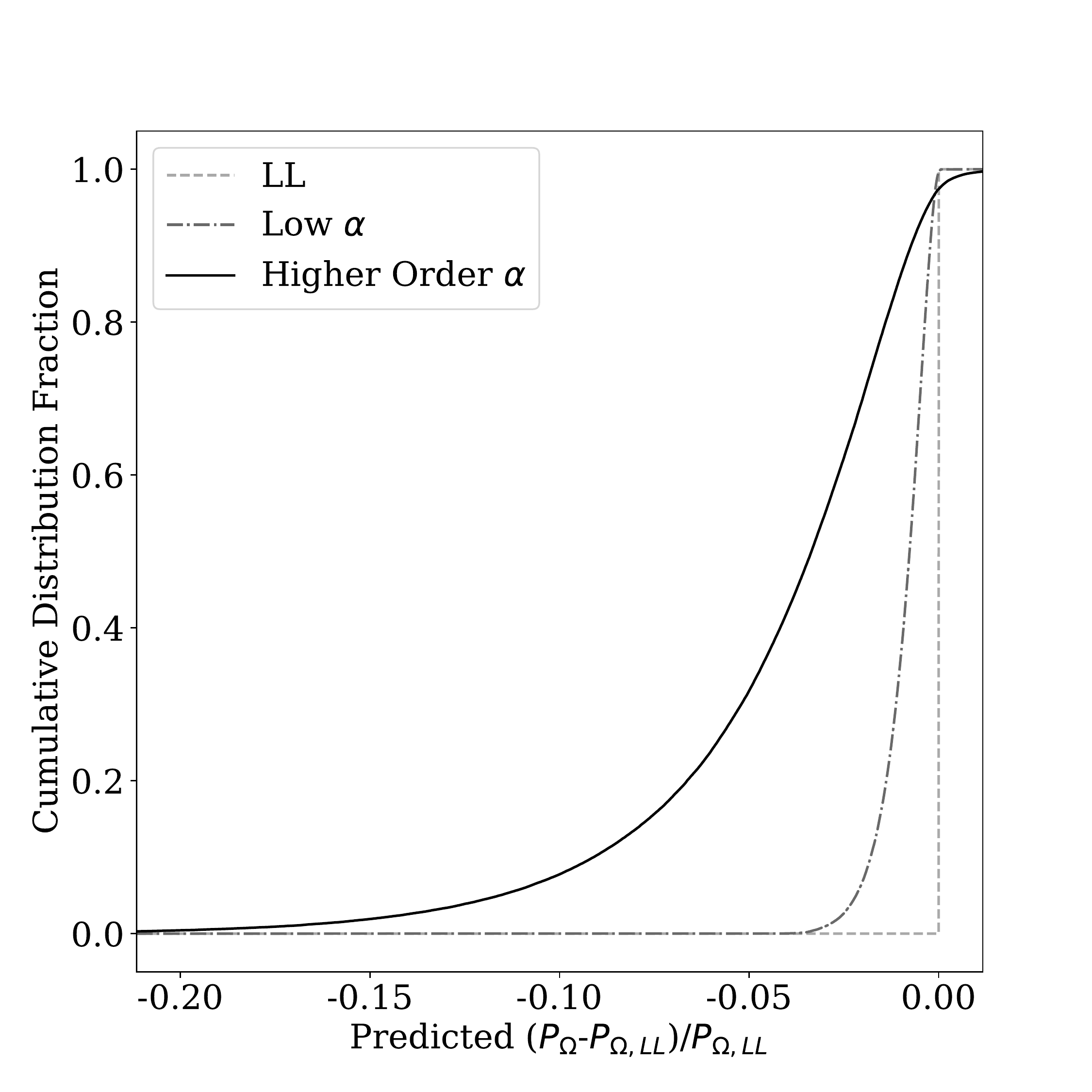}}
  {\caption{Cumulative distribution of expected deviation of nodal precession period from the linear theory prediction for the sample of \emph{Kepler}-like multiplanet systems for our model (solid black) and the $\alpha\ll1$ prediction from \cite{2011Lithwick} (dash-dot gray). The reference linear theory at zero is shown in dashed gray. The x-axis limits are the 0.3 and 99.7 percentiles of the fourth-order model distribution.}}
  \label{fig:Kdevs}
\end{figure}

The resulting expected deviations from the linear theory prediction are shown in Figure~\ref{fig:Kdevs}, along with the predictions from the $\alpha\ll1$ case from \cite{2011Lithwick}. To calculate the distribution, only sets of $e$, $i$, and $\alpha$ where the combined eccentricity is $<$0.27 and the mutual inclination is $<$15$^\circ$ (where the analytical solution is expected to be accurate) are used, and the median nodal period from the 100 sub-samples for each set are used to calculate the distribution. The spread of deviations is much wider when the higher order effects are taken into account using our solution, varying up to tens of percent and with \textit{approximately 50\% having greater than 5\% deviation from linear theory}. Therefore, for precision work on realistic planetary systems, a higher-order theory than linear Laplace-Lagrange theory is essential.

\section{Secular Resonance}\label{sec:secularres}

For a test particle in a system with two planets, it will experience a large forced inclination at locations where its secular frequency is equal to the planets' nodal precession frequency. In Laplace-Lagrange theory, the test particle's secular frequency can be calculated as:

\begin{equation}
    B = -n\frac{1}{4}\sum_{j=1}^2\frac{m_j}{M_\star}\alpha_j \bar{\alpha_j} b_{3/2}^{(1)}(\alpha_j)
\end{equation}

\citep[Equation 7.57]{MurrayDermott}. This frequency is equal to the sum of the linear theory frequencies of each test particle-planet pair. Replacing the linear theory frequency instead with the fourth-order frequency calculated from Equations~\ref{eqn:Omdot1}/\ref{eqn:Omdot2}, we can calculate how the expected location of the secular resonance will change.

\begin{figure*}[h]
\centering
  \includegraphics[width=\linewidth]{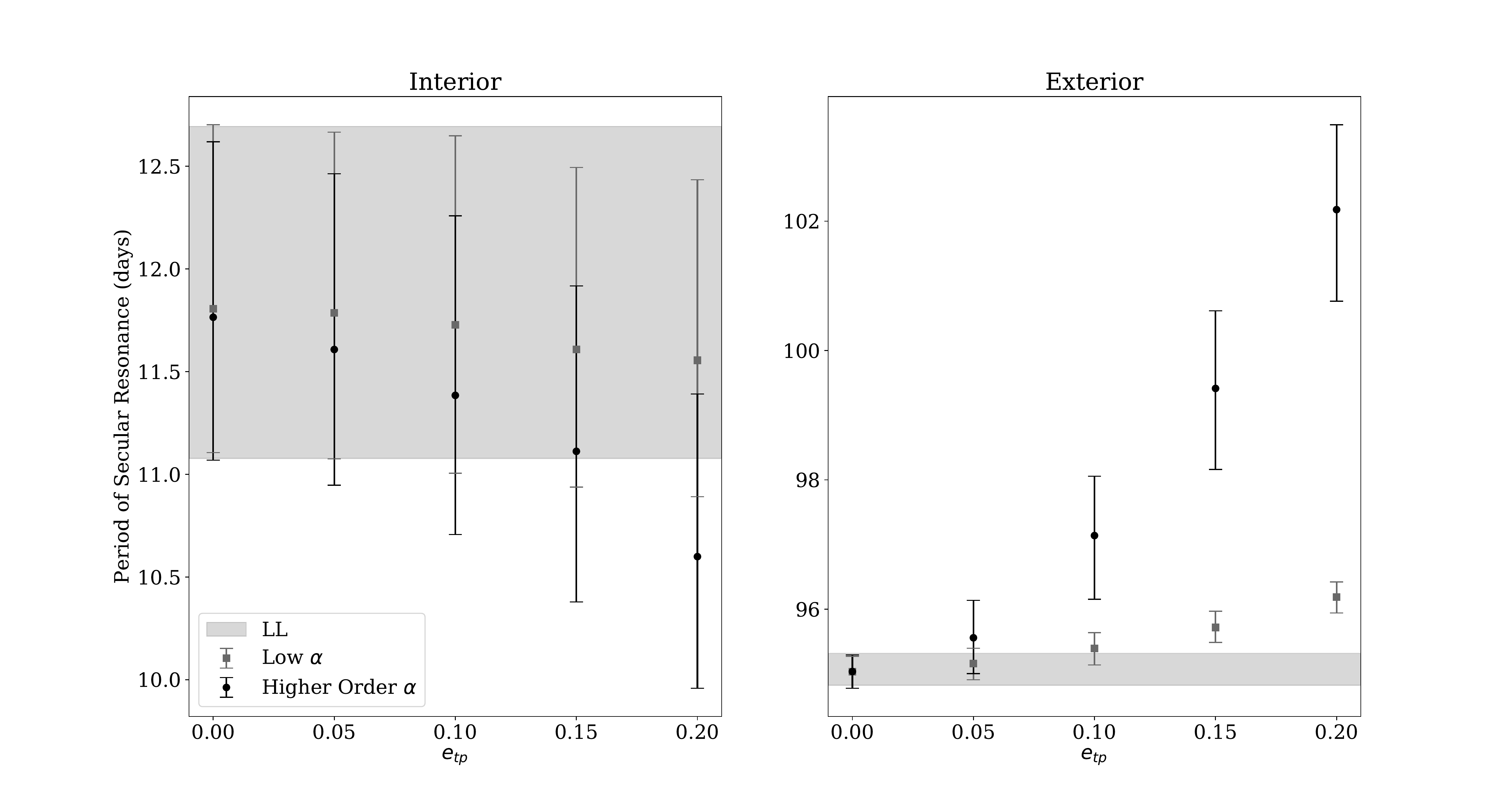}
  \caption{Predicted location of the secular resonance locations of a test particle in the \emph{Kepler}-117 system. Error bars show the 68\% interval around the median value.}
  \label{fig:K117_res}
\end{figure*}

\begin{table}[]
\centering
\caption{\emph{Kepler}-117 system parameters from \cite{2015Bruno}. Angles marked with an asterisk are referenced to the invariable plane, assuming $\Omega_b=\Omega_c=0$ in the sky plane and $\Delta \Omega$ = 180$^\circ$ in the invariable plane.}
\label{tab:K117_param}
\begin{tabular}{|l|l|l|}
\hline
Stellar Mass & \multicolumn{2}{l|}{1.129$^{+0.13}_{-0.023}$ $M_\odot$} \\ \hline
\textbf{Planet:} & \textbf{b} & \textbf{c} \\ \hline
$P$ (days) & 18.7959228(75) & 50.790391(14) \\ \hline
$a$ (AU) & 0.1445$^{+0.0047}_{-0.0014}$ & 0.2804$^{+0.0014}_{-0.0028}$ \\ \hline
$m$ ($M_J$) & 0.094 $\pm$ 0.033 & 1.84 $\pm$ 0.18 \\ \hline
$e$ & 0.0493 $\pm$ 0.0062 & 0.0323 $\pm$ 0.0033 \\ \hline
\multirow{2}{*}{$\omega$ (deg)} & 254.3 $\pm$ 4.1 & 305.0 $\pm$ 7.5 \\ \cline{2-3} 
 & 74.3 $\pm$ 4.1 * & 304.6 $\pm$ 11.4 * \\ \hline
\multirow{2}{*}{$i$ (deg)} & 88.74 $\pm$ 0.12 & 89.64 $\pm$ 0.10 \\ \cline{2-3} 
 & 0.87 $\pm$ 0.16 * & 0.032 $\pm$ 0.01 * \\ \hline
\end{tabular}
\end{table}

To examine the secular resonance locations, we looked at the \emph{Kepler}-117 system. This is a 2-planet system at a low period ratio, not in mean-motion resonance, with known masses and orbits. The planetary properties and stellar mass are taken from \citet[Table 3]{2015Bruno} and summarized here in Table~\ref{tab:K117_param}. The system is rotated into the invariable plane for each calculation.

We account for observational uncertainties by making a sample of 1000 sets of system properties drawn from normal distributions based on the published best fit values and uncertainties. We also account for the possible orientation of the test particle by sampling for 30 $\omega$ values. The test particle is placed in the system's invariable plane, and the linear, second-order, and fourth-order solutions are calculated with various eccentricities of the test particle. Results are shown in Figure~\ref{fig:K117_res}. 

For the interior resonance, observational uncertainty is large enough that the higher order solutions are not separated from the linear theory solution. The uncertainty for this resonance is dominated by the uncertainty in the mass measurement of \emph{Kepler}-117 b, and improved observations could result in differentiation between the models. For the exterior resonance, the difference in location in the higher order solutions is significant from that of linear theory. The uncertainty for this resonance is dominated by the hypothetical orientation ($\omega$) of the test particle and not the observational uncertainties.

Because of the large inclination expected for a test particle in secular resonance, accurately understanding the location of the secular resonance can help predict where planets might be expected to be transit. The dependence on eccentricity could provide a constraint on the eccentricity of a transiting planet near the resonance location.

Additionally, secular resonances within a system, both in the nodal and apsidal frequencies, can interact and lead to chaos when they overlap \citep{2011Lithwick,2014Lithwick}. These overlapping secular resonances could lead to systems maintaining marginal stability over their lifetimes; for example, Mercury in the present-day Solar System \citep{1996Laskar}. Understanding the impact of higher-order effects on the location and width of secular resonances could lead to a better understanding and prediction of exoplanet system architectures.

\section{Conclusions}\label{sec:concl}

We used numerical and analytical methods to probe the nodal precession rate of exoplanet systems. From our simulated systems, we find that the second-order nonlinear nodal precession rate given by \cite{2011Lithwick} is a good approximation when $\alpha~\ll~1$. However, the sensitivity of this precession rate to inclination and eccentricity is a factor of several times more for period ratios relevant to the planetary systems observed by \emph{Kepler} (see Section~\ref{sec:Kepler}). Inclination values of a few degrees and eccentricity values of a few percent are expected to change the nodal rates by a few or even tens of percent. Given that the dependence is on the square of these quantities, the effect would be even more significant at higher values of inclination and eccentricity.

We present a higher-order nonlinear solution for the nodal precession rate, using \cite{MurrayDermott}'s expansion of the disturbing function to fourth order in eccentricity and inclination, maintaining all associated $\alpha$ terms. We find that this solution is a good descriptor of nodal precession rates for simulated systems with moderate eccentricity and inclination at a wide range of period ratios (see Section~\ref{sec:applicablerange}).

The locations and widths of secular resonances in \emph{Kepler}-like systems could deviate significantly from that expected from linear theory, as seen in the expected locations of the inclination-node secular resonances for an eccentric test particle in the \emph{Kepler}-117 system. Further, the overlap of secular resonances can lead to secular chaos in exoplanet systems, which could drive architecture ordering as systems maintain marginal stability on the edge of the overlapping secular resonances. Understanding the nonlinearity of precession rates is critical to understanding the interaction of secular resonances and their effect on exoplanet system architectures.

\acknowledgments

We acknowledge support of grant NASA-NNX17AB93G through NASA's Exoplanet Research Program. We thank Eric Ford and Yoram Lithwick for helpful discussions and Gwena\"el Bou\'e and the anonymous referee for comments on a draft of the manuscript.

\software{\texttt{REBOUND} (http://github.com/hannorein/rebound)}

\appendix

\section{Expressions of \texorpdfstring{$\lowercase{f}(\alpha)$}{f(alpha)}}\label{apsec:ffuncs}

These expressions make use of of the Laplace coefficients $b_s^{(j)}$, defined as

\begin{equation}\label{eqn:laplacecoeff}
    b_s^{(j)} (\alpha) = \frac{1}{\pi} \int_{0}^{2\pi} \frac{cos\, j\psi\, \mathrm{d}\psi}{(1-2\alpha cos\psi + \alpha^2)^s}.
\end{equation}

\begin{equation}\label{eqn:f3}
    f_3 = -\tfrac{1}{2}\alpha b_{3/2}^{(1)}
\end{equation}

\begin{equation}\label{eqn:f7}
    f_7 = \tfrac{1}{16} \left [-4\alpha b_{3/2}^{(1)} -12\alpha^2(b_{5/2}^{(0)}+b_{5/2}^{(2)})+30\alpha^3 b_{5/2}^{(1)} - \tfrac{15}{2} \alpha^3(b_{7/2}^{(3)}-4\alpha b_{7/2}^{(2)}+(4\alpha^2+3)b_{7/2}^{(1)} -4\alpha b_{7/2}^{(0)}) \right ]
\end{equation}

\begin{equation}\label{eqn:f8}
    f_8 = \tfrac{3}{4} \alpha^2 \left (\tfrac{1}{2} b_{5/2}^{(2)} + b_{5/2}^{(0)} \right)
\end{equation}

\begin{equation}\label{eqn:f9}
    f_9 = \tfrac{1}{4}\alpha \left (2b_{3/2}^{(1)} + 3\alpha b_{5/2}^{(2)} + 15\alpha b_{5/2}^{(0)} \right )
\end{equation}

\begin{equation}\label{eqn:f13}
    f_{13} = \tfrac{1}{8} \left[6 \alpha^2 (b_{5/2}^{(3)}+3 b_{5/2}^{(1)})-15 \alpha^3 (b_{5/2}^{(2)}+b_{5/2}^{(0)})+\tfrac{15}{4} \alpha^3 (b_{7/2}^{(4)}-4 \alpha b_{7/2}^{(3)}+(4 \alpha^2+4) b_{7/2}^{(2)} - 12 \alpha b_{7/2}^{(1)}+(4 \alpha^2+3) b_{7/2}^{(0)}) \right]
\end{equation}

\begin{equation}\label{eqn:f14}
    f_{14} = \alpha b_{3/2}^{(1)}
\end{equation}

\begin{equation}\label{eqn:f15}
    f_{15} = \tfrac{1}{4} \left[2 \alpha b_{3/2}^{(1)}+ 6 \alpha^2 (b_{5/2}^{(0)}+b_{5/2}^{(2)}) -15 \alpha^3 b_{5/2}^{(1)} + \tfrac{15}{4} \alpha^3 \left(b_{7/2}^{(3)}-4 \alpha b_{7/2}^{(2)}+(4 \alpha^2+3) b_{7/2}^{(1)}-4 \alpha b_{7/2}^{(0)}\right) \right]
\end{equation}

\begin{equation}\label{eqn:f16}
    f_{16} = -\alpha \left(\tfrac{1}{2} b_{3/2}^{(1)}+3 \alpha b_{5/2}^{(0)}+ \tfrac{3}{2} \alpha b_{5/2}^{(2)} \right)
\end{equation}

\begin{equation}\label{eqn:f18}
    f_{18} = \tfrac{1}{16} \left [12 \alpha b_{3/2}^{(1)} +12 \alpha^2 (b_{5/2}^{(2)}+b_{5/2}^{(0)}) -27 \alpha^3 b_{5/2}^{(1)} + \tfrac{15}{4} \alpha^3 \left(b_{7/2}^{(3)}-4 \alpha b_{7/2}^{(2)}+(4 \alpha^2+3) b_{7/2}^{(1)}-4 \alpha b_{7/2}^{(0)} \right)  \right ]
\end{equation}

\begin{equation}\label{eqn:f19}
    f_{19} = \tfrac{1}{8} \left[ -12 \alpha^2 b_{5/2}^{(1)}+15 \alpha^3 b_{5/2}^{(2)}-
               \tfrac{15}{4} \alpha^3 \left (2 b_{7/2}^{(2)}-8 \alpha b_{7/2}^{(1)}+(4 \alpha^2+2) b_{7/2}^{(0)} \right) \right ]
\end{equation}

\begin{equation}\label{eqn:f20}
    f_{20} = \tfrac{1}{16} \left[ \tfrac{15}{4} \alpha^3 \left( b_{7/2}^{(3)}-4 \alpha b_{7/2}^{(2)}+(4 \alpha^2+3) b_{7/2}^{(1)} - 4 \alpha b_{7/2}^{(0)}-4/5 b_{5/2}^{(1)} \right)  \right ]
\end{equation}

\begin{equation}\label{eqn:f21}
    f_{21} = -2f_{18}
\end{equation}

\begin{equation}\label{eqn:f22}
    f_{22} = 2f_{19}
\end{equation}

\begin{equation}\label{eqn:f23}
    f_{23} = \tfrac{1}{4} \left[-6 \alpha^2 (b_{5/2}^{(1)}+b_{5/2}^{(3)})+15 \alpha^3 b_{5/2}^{(2)}- \tfrac{15}{4} \alpha^3 \left(b_{7/2}^{(4)}-4 \alpha b_{7/2}^{(3)}+(4 \alpha^2+2) b_{7/2}^{(2)} -4 \alpha b_{7/2}^{(1)}+b_{7/2}^{(0)}\right) \right]
\end{equation}

\begin{equation}\label{eqn:f24}
    f_{24} = -2f_{19}
\end{equation}

\begin{equation}\label{eqn:f25}
    f_{25} = -2f_{20}
\end{equation}

\begin{equation}\label{eqn:f26}
    f_{26} = \tfrac{1}{2} \alpha \left( b_{3/2}^{(1)}+ \tfrac{3}{2} \alpha b_{5/2}^{(0)}+3 \alpha b_{5/2}^{(2)} \right)
\end{equation}

\newpage

\section{Tabulated Results}\label{appsec:full_tables}

\begin{splitdeluxetable}{cccccccccccBcccccc}
\tablecaption{Restricted Simulations Data}
\label{apptab:restricted_results}
\tabletypesize{\scriptsize}
\tablehead{
\colhead{$P_1$}  & \colhead{$P_2$} & \colhead{Period Ratio}
& \colhead{$e_{1,0}$} & \colhead{$e_{2,0}$} & \colhead{$i_{1,0}$}
& \colhead{$i_{2,0}$} & \colhead{$\omega_{1,0}$} 
& \colhead{$\omega_{2,0}$}  & \colhead{$P_{\Omega,2}$}
 & \colhead{$\sigma_{P_{\Omega,2}}$}  & \colhead{$t_\text{stable}$}
  & \colhead{$t_\text{sim}$}  & \colhead{$P_{\Omega,\text{LL}}$}
   & \colhead{$P_{\Omega,\text{2nd}}$} & \colhead{$P_{\Omega,\text{4th}}$}  & \colhead{Figures Used} \\
\colhead{(d)}  & \colhead{(d)} & \colhead{}
& \colhead{} & \colhead{} & \colhead{(rad)}
& \colhead{(rad)} & \colhead{(rad)} 
& \colhead{(rad)}  & \colhead{(d)}
 & \colhead{(d)}  & \colhead{(d)}
  & \colhead{(d)}  & \colhead{(d)}
   & \colhead{(d)} & \colhead{(d)}  & \colhead{}
} 
\startdata
3.48 & 6.17 & 1.77 & 2.090E-16 & 4.49E-17 & 0 & 0.0407 & 0.827 & 1.094 & 122124.51 & 0.00238 & 616559.81 & 616559.81 & 121025.25 & 121125.70 & 122162.63 & 1,3 \\
3.48 & 6.32 & 1.816 & 2.090E-16 & 0.033 & 0 & 0.0457 & 0.827 & 1.094 & 135455.84 & 0.01463 & 632497.09 & 632497.09 & 135768.52 & 135834.68 & 135521.25 & 3 \\
3.48 & 7.33 & 2.105 & 2.090E-16 & 6.33E-17 & 0 & 0.0696 & 0.827 & 1.094 & 253685.21 & 0.03683 & 109987.50 & 109987.50 & 249262.47 & 249867.71 & 253485.62 & 1,3 \\
3.48 & 7.67 & 2.201 & 2.090E-16 & 0.085 & 0 & 0.1047 & 0.827 & 1.094 & 290324.28 & 0.16596 & 766868.72 & 766868.72 & 295058.11 & 295605.46 & 289729.69 & 3 \\
3.48 & 8.15 & 2.339 & 2.090E-16 & 0.100 & 0 & 0.0457 & 0.827 & 1.094 & 345870.47 & 0.30217 & 814679.14 & 814679.14 & 367177.73 & 365730.40 & 346743.21 & 3 \\
3.48 & 9.26 & 2.659 & 2.090E-16 & 0.067 & 0 & 0.0087 & 0.827 & 1.094 & 555062.52 & 0.17255 & 926234.47 & 926234.47 & 568096.28 & 566858.18 & 555203.21 & 1,3 \\
3.48 & 11.15 & 3.201 & 2.090E-16 & 0.067 & 0 & 0.0457 & 0.827 & 1.094 & 1001488.27 & 0.26432 & 1115209.17 & 1115209.17 & 1016797.08 & 1015599.07 & 1001465.71 & 3 \\
3.48 & 11.47 & 3.293 & 2.090E-16 & 0.085 & 0 & 0.0407 & 0.827 & 1.094 & 1076858.79 & 0.43548 & 1147082.32 & 1147082.32 & 1106370.23 & 1103278.56 & 1077204.18 & 3 \\
3.48 & 26.20 & 7.521 & 2.090E-16 & 0.033 & 0 & 0.0727 & 0.827 & 1.094 & 10254247.81 & 0.33899 & 2619938.20 & 2619938.20 & 10240518.29 & 10261952.35 & 10252821.81 & 3 \\
3.48 & 174.17 & 50 & 2.090E-16 & 0.100 & 0 & 0.1047 & 0.827 & 1.094 & 947110434.65 & 1.01768 & 17417019.80 & 17417019.80 & 961145943.34 & 961610510.41 & 947155267.41 & 3 \\
\enddata
\tablecomments{Table~\ref{apptab:restricted_results} is published in its entirety in the machine-readable format. A random subset of rows are shown here for guidance regarding its form and content.}
\end{splitdeluxetable}

\begin{splitdeluxetable}{cccccccccccccBcccccc}
\tablecaption{Unrestricted Simulations Data}
\label{apptab:unrestricted_results}
\tabletypesize{\scriptsize}
\tablehead{
\colhead{$P_1$}  & \colhead{$P_2$} & \colhead{Period Ratio}
& \colhead{$e_{1,0}$} & \colhead{$e_{2,0}$} & \colhead{$i_{1,0}$}
& \colhead{$i_{2,0}$} & \colhead{$\omega_{1,0}$} 
& \colhead{$\omega_{2,0}$} & \colhead{$P_{\Omega,1}$}
& \colhead{$\sigma_{P_{\Omega,1}}$} & \colhead{$P_{\Omega,2}$}
 & \colhead{$\sigma_{P_{\Omega,2}}$}  & \colhead{$t_\text{stable}$}
  & \colhead{$t_\text{sim}$}  & \colhead{$P_{\Omega,\text{LL}}$}
   & \colhead{$P_{\Omega,\text{2nd}}$} & \colhead{$P_{\Omega,\text{4th}}$}  & \colhead{Figures Used} \\
\colhead{(d)}  & \colhead{(d)} & \colhead{}
& \colhead{} & \colhead{} & \colhead{(rad)}
& \colhead{(rad)} & \colhead{(rad)} 
& \colhead{(rad)}  & \colhead{(d)}
& \colhead{(d)}  & \colhead{(d)}
 & \colhead{(d)}  & \colhead{(d)}
  & \colhead{(d)}  & \colhead{(d)}
   & \colhead{(d)} & \colhead{(d)}  & \colhead{}
} 
\startdata
6.24 & 11.05 & 1.77 & 2.95E-07 & 3.57E-07 & 0.00497 & 0.00376 & -1.202 & -1.489 & 141663.88 & 0.00276 & 141663.88 & 0.00276 & 1105326.61 & 1105326.61 & 141883.75 & 141889.15 & 141945.60 & 2,4,6,7 \\
6.24 & 16.89 & 2.705 & 0.112 & 0.112 & 0.11861 & 0.07770 & 2.556 & 2.144 & 577689.39 & 3.92005 & 577691.82 & 3.92102 & 1689057.79 & 1689057.79 & 648060.03 & 644268.54 & 573627.92 & 6,7 \\
6.24 & 16.89 & 2.705 & 0.149 & 0.149 & 0.19252 & 0.12583 & 2.558 & 2.146 & 584151.97 & 7.40156 & 584158.33 & 7.39694 & 1689057.79 & 1689057.79 & 648060.03 & 652144.30 & 560039.82 & 6,7 \\
6.24 & 18.03 & 2.888 & 1.99E-05 & 2.40E-05 & 0.02781 & 0.01785 & -1.202 & -1.489 & 791773.06 & 0.01935 & 791773.17 & 0.01936 & 1803337.58 & 1803337.58 & 788337.06 & 789159.83 & 791589.78 & 2,4,6,7 \\
6.24 & 19.99 & 3.201 & 0.099 & 0.099 & 0.15899 & 0.09833 & 2.558 & 2.147 & 1049403.61 & 6.96787 & 1049428.86 & 6.98544 & 1999266.84 & 1999266.84 & 1060123.12 & 1074537.96 & 1041708.09 & 6,7 \\
6.24 & 20.85 & 3.339 & 8.85E-07 & 1.07E-06 & 0.00541 & 0.00331 & -1.202 & -1.489 & 1191726.67 & 0.00915 & 1191726.70 & 0.00915 & 2084977.93 & 2084977.93 & 1191449.91 & 1191495.27 & 1191594.60 & 2,4,6,7 \\
6.24 & 21.71 & 3.476 & 0.200 & 0.200 & 0.02848 & 0.01718 & 2.554 & 2.142 & 815342.02 & 14.11037 & 815342.06 & 14.11079 & 2170686.53 & 2170686.53 & 1330353.71 & 1233025.66 & 895405.98 & 6,7 \\
6.24 & 24.85 & 3.979 & 0.112 & 0.112 & 0.12457 & 0.07172 & 2.557 & 2.145 & 1769718.56 & 16.38353 & 1769718.22 & 16.41041 & 2484957.20 & 2484957.20 & 1904627.02 & 1893673.45 & 1765846.63 & 6,7 \\
6.24 & 64.41 & 10.313 & 0.084 & 0.084 & 0.11679 & 0.04892 & 2.560 & 2.149 & 17709895.06 & 73.78958 & 17709940.33 & 74.17984 & 93574880.08 & 93574880.08 & 18050503.76 & 18041931.10 & 17708250.30 & 6,7 \\
6.24 & 121.10 & 19.392 & 8.04E-05 & 9.72E-05 & 0.03039 & 0.01034 & -1.202 & -1.489 & 71090700.91 & 0.25767 & 71090774.55 & 0.75852 & 361761267.20 & 361761267.20 & 71020986.85 & 71079921.86 & 71097363.14 & 2,4,6,7 \\
6.24 & 13.18 & 2.11 & 0.900 & 0.900 & 0.02378 & 0.01695 & 2.554 & 2.142 & nan & nan & nan & nan & 2945.05 & 1317649.72 & 285089.41 & 108849.70 & 5453.29 &  \\
6.24 & 312.24 & 50 & 0.066 & 0.066 & 0.08389 & 0.02077 & 2.561 & 2.149 & 514094177.93 & 142.50032 & 514102083.46 & 146.94138 & 2547874656.26 & 2547874656.26 & 520413164.36 & 518753723.38 & 514589013.77 & 4,6,7 \\
\enddata
\tablecomments{Table~\ref{apptab:unrestricted_results} is published in its entirety in the machine-readable format. A random subset of rows are shown here for guidance regarding its form and content. A nan value for the simulated precession period indicates that a sinusoid could not be fit to the $\Omega$ data, usually in the case of non-secularly stable systems.}
\end{splitdeluxetable}

\newpage
\bibliography{NPPRbib}
\bibliographystyle{aasjournal}

\end{document}